\documentclass[a4paper]{article}
\usepackage{amsmath,amsfonts}
\usepackage{algorithmic}
\usepackage{algorithm}
\usepackage{array}
\usepackage[caption=false,font=normalsize,labelfont=up,textfont=up]{subfig}
\usepackage{textcomp}
\usepackage{stfloats}
\usepackage{verbatim}
\usepackage{graphicx}
\usepackage{cite}
\usepackage{multirow}
\usepackage{subfig}
\usepackage{xcolor}
\usepackage{verbatim}
\usepackage{booktabs}
\usepackage{arydshln}
\usepackage{bbding}
\usepackage{url}
\usepackage[hidelinks]{hyperref}

\usepackage{INTERSPEECH2022}

\title{DRAFT: A Novel Framework to Reduce Domain Shifting in Self-supervised Learning and Its Application to Children's ASR}
\name{Ruchao Fan$^1$\thanks{This paper was supported in part by the NSF and the UCLA-Amazon Science Hub.}, Abeer Alwan$^1$}
\address{
  $^1$Dept. of Electrical and Computer Engineering, University of California, Los Angeles, USA} 
\email{fanruchao@g.ucla.edu, alwan@ee.ucla.edu}

\begin{document}

\maketitle
\begin{abstract}
Self-supervised learning (SSL) in the pretraining stage using un-annotated speech data has been successful in low-resource automatic speech recognition (ASR) tasks. However, models trained through SSL are biased to the pretraining data which is usually different from the data used in finetuning tasks, causing a domain shifting problem, and thus resulting in limited knowledge transfer. We propose a novel framework, domain responsible adaptation and finetuning (DRAFT), to reduce domain shifting in pretrained speech models through an additional adaptation stage. In DRAFT, residual adapters (RAs) are inserted in the pretrained model to learn domain-related information with the same SSL loss as the pretraining stage. Only RA parameters are updated during the adaptation stage. DRAFT is agnostic to the type of SSL method used and is evaluated with three widely used approaches: APC, Wav2vec2.0, and HuBERT. On two child ASR tasks (OGI and MyST databases), using SSL models trained with un-annotated adult speech data (Librispeech), relative WER improvements of up to 19.7\% are observed when compared to the pretrained models without adaptation. Additional experiments examined the potential of cross knowledge transfer between the two datasets and the results are promising, showing a broader usage of the proposed DRAFT framework.

\end{abstract}

\section{Introduction}



Recently, self-supervised learning (SSL) for speech has been investigated\cite{chen2021wavlm,zhang2021bigssl,chung2021w2v,ao2021speecht5,liu2021tera} because of its great potential in improving low-resource ASR tasks. In SSL, pseudo-labels are generated for un-annotated data for model pretraining, and then the learned knowledge is transferred to a downstream supervised task through finetuning. For example, autoregressive predictive coding (APC) uses a shifted input sequence as supervision such that the model predicts future frames from previous frames\cite{chung2019unsupervised,chung2020generative,Ravi2020}. Different from APC, Wav2vec-based methods include sampled negative frames in a contrastive loss to increase discrimination between frames in a way that the learned embedding is closer to the positive frame and more distant to the negative frames\cite{oord2018representation,schneider2019wav2vec,baevski2020wav2vec,baevski2019vq}. A more recent SSL framework, HuBERT\cite{hsu2021hubertc,hsu2021hubertj}, creates a pseudo-label for each speech frame using clustering techniques like K-means. Models learned with SSL objectives can be used in two manners: 1) feature extraction as a replacement of hand-crafted speech features\cite{yang2021superb,evain2021lebenchmark,chang2021exploration}; or 2) model initialization for finetuning downstream tasks\cite{jiang2021further,wang2020unsupervised,misra2021comparison}. SSL has been shown to be effective in ASR of low-resource languages\cite{riviere2020unsupervised,yi2020applying}, noisy speech\cite{wang2021wav2vec}, accented speech\cite{li2021accent}, and child ASR\cite{fan2021bi,wang2021low}. In \cite{fan2021bi}, we extend APC to learn bidirectional contexts for pretraining from adult speech data.
 
However, a main weakness of SSL is that training from one domain causes domain shifting when finetuning on data from a different domain\cite{meng2021don,sanabria2022measuring}. To address this issue, previous work presented robust pretrained models by adding target domain data during pretraining\cite{hsu2021robust,hwang2021large}. However, including target domain data might not be feasible at the pretraining stage. In addition, retraining a large-scale SSL model with both the source and target domain data may not be computationally efficient. In \cite{khurana2021magic}, un-annotated target domain data are used for semi-supervised learning during the finetuning stage to alleviate the performance degradation caused by domain shifting. However, no previous work, to our knowledge, has investigated methods for performing adaptation of self-supervised models with finetuning data.

In this paper, we propose a novel framework, domain responsible adaptation and finetuning (DRAFT), to reduce domain shifting in SSL-pretrained speech models. In DRAFT, residual adapters (RAs) are placed between blocks in the transformer and are responsible for learning domain specific information during an additional adaptation stage. The additional adaptation stage trains the model with finetuning data and with the same SSL loss that was used in the pretraining stage. To prevent catastrophic forgetting of the learned knowledge from source domain data, only RA parameters are updated during the adaptation stage. Hence, DRAFT has a lower computational cost than retraining the pretrained models with both the source and target domain data. Note that DRAFT is universal to different SSL methods. When performing DRAFT on SSL-pretrained speech models (trained with adult speech data) for child ASR tasks, we obtain significant improvements over baselines without adaptation for both causal (pretrained with APC) and non-causal transformers (pretrained with Wav2vec2.0 or HuBERT). We presented partial results of DRAFT in a paper that is in review\cite{fan2022towards}. In this paper, more and different experiments are presented. For example, when the learned RAs from one task are used for finetuning another task (cross transfer), improvements are still observed, showing a broader usage of the proposed framework. We also include results of adapter finetuning experiments for more detailed comparisons.

Note that residual adapters have been used before to learn domain specific parameters for adaptation\cite{hou2021exploiting,kannan2019large,rebuffi2017learning, tomanek2021residual}. In \cite{houlsby2019parameter} residual adapters are inserted to achieve a parameter efficient adaptation for disordered speech. In \cite{thomas2022efficient}, the same idea is used for SSL-pretrained models for efficient adaptation, which is known as adapter tuning in natural language processing. However, these methods use residual adapters for supervised tasks or during finetuning, unlike our proposed method. 

The remainder of this paper is organized as follows. Section \ref{sec:method} describes the proposed DRAFT framework. Experimental setups are described in Section \ref{sec:exp_setup}. Results are shown and discussed in Section \ref{sec:result}. We conclude the paper in Section \ref{sec:conclusion}.


\section{Method}
\label{sec:method}

\begin{figure*}[th]
\centering
\centerline{\includegraphics[width=0.98\textwidth,height=0.36\textwidth]{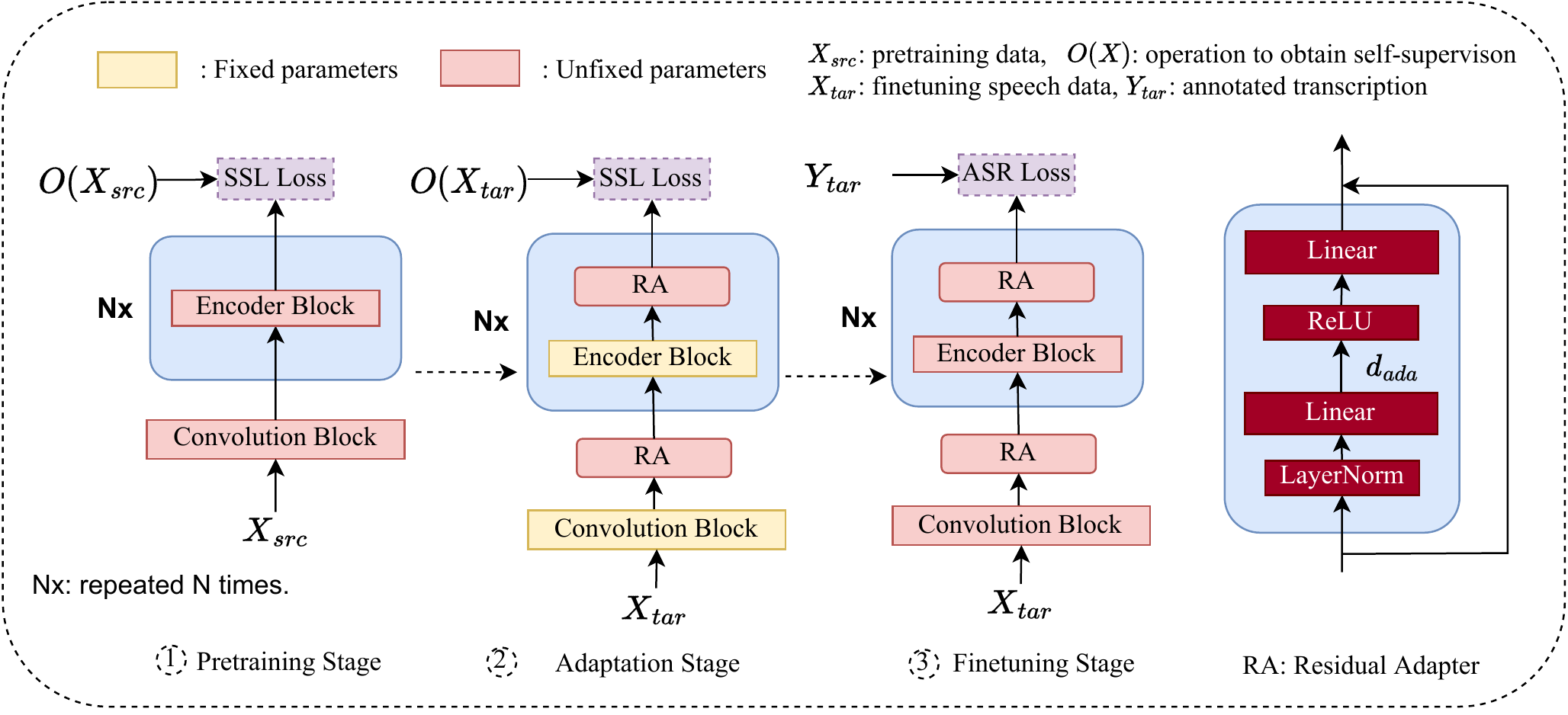}}
\caption{An overview of DRAFT. $d_{ada}$ is the output dimension of the first linear layer in  the residual adapter (RA).}
\label{fig:draft}
\end{figure*}

In this section, we introduce the proposed framework: domain responsible adaptation and finetuning (DRAFT). Fig.\ref{fig:draft} shows an overview of DRAFT and the structure of the residual adapters.

\subsection{Residual Adapters}
\label{ssec:res_ada}
We start with a description of residual adapters (RAs) (shown in the right side of Fig.\ref{fig:draft}) since they are the most important modules in DRAFT. Specifically, an RA consists of two feed-forward layers with a residual connection. The activation function between the two feed-forward layers makes the adapter non-linear. A layer normalization is inserted at the beginning, which is similar to a self-attention block. We define the dimension of the first feed-forward layer output as $d_{ada}$, which determines the number of parameters of the RA. The effect of $d_{ada}$ on performance will be analysed in Sec.\ref{ssec:result_effect_ada}. Note that RA can be placed anywhere in the model. In our case, we insert one residual adapter after the convolution block and one after each encoder block as shown in Fig.\ref{fig:draft}. We assume that the output of each block needs to be transformed to be similar to that of the target domain data so that the model can easily converge.

\subsection{Reducing Domain Shifting in SSL-pretrained Models}
\label{ssec:draft}
Learning from self-supervised pretrained models as a starting point is helpful because of the benefits of learning from large-amounts of un-annotated data. However, performance improvement could potentially be limited due to the domain mismatch between the pretraining and finetuning data. As mentioned in the Introduction, including data from the target domain in the pretraining stage can further improve the performance of the target task, but it requires re-training of the self-supervised model, which is time-consuming and computationally expensive. It would be more practical to adapt the pretrained models with only target data. However, direct adaptation on self-supervised models may lead to a catastrophic forgetting problem. We therefore propose DRAFT, a domain responsible adaptation and finetuning framework to prevent catastrophic forgetting that happens when finetuning the entire pretrained model, and to address the domain shifting problem of the conventional self-supervised pretraining and finetuning paradigm.
\subsubsection{Simple Adaptation and Finetuning (SAFT)}
\label{sssec:saft}
Before introducing DRAFT, we would like to discuss the most direct way of doing adaptation by re-training the pretrained models, which we refer to as simple adaptation and finetuning (SAFT). In SAFT, an adaptation stage is inserted between the pretraining and finetuning stage, and the model is adapted with an SSL loss and with the finetuning data. All model parameters are updated at the adaptation stage with a smaller learning rate than the pretraining stage to prevent overfitting. The model after the adaptation stage is used as initialization for the finetuning stage with an ASR loss. SAFT updates the parameters of the entire model, and thus might have catastrophic forgetting of learned knowledge from the pretraining stage\cite{french1999catastrophic}. 

\subsubsection{Domain Responsible Adapters for Finetuning (DRAFT)}
\label{sssec:res_adapt}
DRAFT is a three-stage (pretraining, adaptation, and finetuning) training paradigm as shown in Fig.\ref{fig:draft}. Different from SAFT, residual adapters are inserted in the model at the adaptation stage to learn knowledge from the finetuning data. Let $\theta_{ada}$ be the parameters in residual adapters, $\theta_f$ the parameters in the backbone model (without residual adapters), $\theta_g$ the parameters in the last embedding mapping layer for the self-supervised task, and $\theta_g'$ the parameters in the last linear layer for the ASR task. Suppose source domain data are $S_{src}$ and target domain data are $S_{tgt}$, DRAFT can be described as:

\textbf{Stage 1:} Initialize a model $\{\theta_{f}^0, \theta_{g}^0\}$, update the parameters using data $S_{src}$ and self-supervised loss $L_{ssl}$, and obtain a pretrained model $\{\theta_{f}^1, \theta_{g}^1\}$.

\textbf{Stage 2:} From model $\{\theta_{f}^1, \theta_{g}^1\}$, insert residual adapters after each block initialized with $\theta_{ada}^0$, freeze $\{\theta_{f}^1, \theta_{g}^1\}$ and update $\theta_{ada}^0$ using data $S_{tgt}$ and the same self-supervised loss $L_{ssl}$, and obtain an adapted model $\{\theta_{f}^1, \theta_{ada}^1, \theta_{g}^1\}$.

\textbf{Stage 3:} From model $\{\theta_{f}^1, \theta_{ada}^1, \theta_{g}^1\}$, replace $\theta_{g}^1$ with a new generator that can map the embedding space to token space as $\theta_{g'}^0$, update the entire model with data $S_{tgt}$ and an ASR loss such as the connectionist temporal classification (CTC) loss function, and obtain the final ASR model $\{\theta_{f}^2, \theta_{ada}^2, \theta_{g'}^1\}$.

Note that the superscript in each $\theta$ is the number of times the parameters are updated. For example, $\theta_f^2$ means that backbone model has been updated twice, once in stage one and the other in stage three. Since the backbone model is frozen during stage 2, catastrophic forgetting is effectively prevented. 

A similar concurrent work to the proposed approach is \cite{kessler2021continual} where residual adapters are also used to prevent catastrophic forgetting, but for continually learning representations from various languages. Different from \cite{kessler2021continual}, finetuning data are used in an additional stage with the purpose of adapting SSL-pretrained models in our method. Furthermore, we update all the parameters during the finetuning stage instead of fixing the backbone model parameters because fixing the parameter does not perform well in our experiments.


\section{Experimental Settings}
\label{sec:exp_setup}
Because of the availability of large databases of adult speech, we explore how SSL methods trained with adult speech can help the development of child ASR systems. 

\subsection{Databases}
\label{ssec:exp_data}
Librispeech adult speech corpus \cite{panayotov2015librispeech} is used at the pretraining stage. It contains 960 hours of read speech.
During finetuning, we target child ASR tasks on two datasets: OGI Kids' Speech Corpus (OGI)~\cite{shobaki2000ogi} and My Science Tutor (MyST)\cite{ward2011my,ward2019my}. For OGI, the scripted part is used, which contains child speech from approximately 100 speakers per grade (from kindergarten to grade 10). The utterances are randomly split into train (70\%), development (15\%) and test (15\%) sets without speaker overlap\cite{fan2021bi}. As a result, nearly 50 hours of child data are used to train the child ASR system. MyST contains 499 hours of speech data with 244,069 utterances of conversational speech between children and a virtual tutor from 1,372 students between third and fifth grades. However, only 42\% of the corpus (240 hours) is annotated for ASR. We use the annotated part of the corpus to verify the effectiveness of our proposed methods. The corpus also contains a development set and test set for evaluation.

\subsection{DRAFT settings}
Based on pilot experiments, we applied speed perturbation\cite{ko2015audio} and SpecAug\cite{park2019specaugment} to both child datasets for better performance. During evaluation, greedy search decoding is used during evaluation for all the experiments. We examine the proposed DRAFT framework on three widely-used SSL approaches: APC, Wav2vec2.0, and HuBERT.

\subsubsection{APC}
\label{ssec:exp_apc}
We use multiple shifted sequences as supervisions to construct a multi-task training objective for APC to learn richer information from pretraining. 80-dimensional log-filter-bank features are extracted as model input without any concatenation or frame skipping. The model consists of a two-layer convolution block with a sub-sampling of four along the time axis, 12 transformer encoder blocks with a causal mask in self-attention computation, and a linear layer for each shifted sequence. The output has 320 dimensions because of the sub-sampling in the convolution block. Adam optimizer is used with a noam-based scheduler, where the noam factor is 5 and the warmup step is 15k. The model is updated for 130k steps with a batch size of 256.

At the adaptation stage, residual adapters (RAs) are inserted into the pretrained model and only the RA parameters are updated. For the OGI data, RAs are updated in 55k steps with a noam factor of 8, and warmup steps of 10k. For the MyST data, RAs are updated in 74k steps with a noam factor 4 and a warmup step of 15k. The batch size is set to 64 for both datasets. We also perform experiments using SAFT with the above configurations but parameters of the entire model are updated.

At the finetuning stage, CTC loss is used. The model is updated in 240k steps with a batch size of 32, a noam factor of 2, and a warmup step of 10k steps for the OGI data. For the MyST data, the model is updated in 340k steps with a batch size of 64, a noam factor of 2, and a warmup step of 15k steps.

\subsubsection{Wav2vec2.0 and HuBERT}
\label{ssec:exp_wav2vec}
For Wav2vec2.0 and HuBERT, we directly use the open-sourced pretrained models in the Fairseq toolkit\cite{ott2019fairseq}. We choose the base model that has about 95M parameters to evaluate the effectiveness of the proposed DRAFT framework. Note that the number of parameters in the APC pretraining models are about 39M. In addition, the pretrained Wav2vec2.0 and HuBERT models are non-causal transformers.

At the adaptation stage, the residual adapters of Wav2vec2.0/HuBERT are updated in 200k/100k steps with learning rate ramping up from 0 to the peak learning rate in 32k/8k steps, and then decays linearly back to 0, where the peak learning rate is 5e-4. The batch size is set to 16.

At the finetuning stage, the model is updated with a batch size of 64 in 40k steps with a multi-step scheduler where the warmup steps are set to 4k. The peak learning rate of 3e-5/7e-5 holds for the next 16k steps, then exponentially decays to the ratio $\lambda$ of the initial learning rate, where $\lambda$ is set to 0.05.

\section{Results and Discussion}
\label{sec:result}

\subsection{Effect of $d_{ada}$ in DRAFT}
\label{ssec:result_effect_ada}
\begin{table}[t]

\scriptsize

\centering
  \caption{WER results of different values of $d_{ada}$ in residual adapters using APC. SAFT is the sample adaptation and finetuning that updates the entire model at the adaptation stage. DRAFT is the proposed domain responsible adaption and finetuning that updates only residual adapters at the adaptation stage. The total number of updated parameters are also shown in absolute and relative values (compared to the SAFT). All DRAFT performance improvements are statistically significant. }

\begin{tabular}{l c cc cc cc}
\hline
\multirow{2}{*}{~} & \multirow{2}{*}{$d_{ada}$} & \multicolumn{2}{c}{OGI} & \multicolumn{2}{c}{MyST} & \multicolumn{2}{c}{Updated Params} \\ 
\cmidrule(r){3-4} \cmidrule(r){5-6} \cmidrule(r){7-8} 

~ & ~ & dev & test & dev & test & total & relative \\ \hline\hline

Baseline & 0 & 5.9 & 7.0 & 36.7 & 36.3 & - & - \\ 

Finetune& 0 & 5.0 & 6.1 & 32.2 & 31.6 & - & - \\

SAFT & 0 & 5.0& 5.9 & 33.4 & 32.9 & 39.2M & - \\ \hline

\multirow{7}{*}{DRAFT} 
 & 64 & 4.9& 5.7 & 31.9 & 31.0 & 0.9M & 2\% \\ 
 & 128 & 4.7& 5.6 & 31.6 & 30.9 & 1.7M & 4\% \\ 
  & 256 & 4.6& 5.3 & 31.1 & 30.4 & 3.4M & 9\% \\ 
   & 512 & 4.4& 5.2 & 30.9 & 30.2 &6.8M & 17\% \\ 
 & 1024 & 4.4& 4.9 & 30.1 & 29.4 & 13.7M & 35\% \\ 
& 2048 & 4.4& 4.9 & 30.0& 29.3 & 27.3M & 70\% \\ \hline

\label{tab:DRAFT_OGI_MyST}
\end{tabular}
\end{table}
We conducted experiments with different values of $d_{ada}$ in the RAs to examine the impact of the number of adapter parameters, because this number influences both WERs and adaptation efficiency. The experiments are conducted on the OGI and MyST datasets using the APC method. Specifically, $d_{ada}$ values are selected from 64 to 2048 and the results are shown in Table \ref{tab:DRAFT_OGI_MyST}. For reference, we also include the results for the baseline, finetuning from APC and SAFT. Both WER results and the number of parameters that need to be updated during training are also shown in the table. We observe that the WER drops when we increase the number of parameters in the RAs. However, the cost is increased training time at the adaptation stage because more parameters need to be updated. For example, even if 2\% of the parameters are updated, the WER can decrease from 5.9\% to 5.7\% on the OGI test set. Hence, the choice of $d_{ada}$ in DRAFT can be adjusted according to scenarios. For example, one can use a small value of $d_{ada}$ to achieve a fast adaptation of the self-supervised model when the computational resources are limited. A large value of $d_{ada}$ can be used to achieve a better performance for the finetuning task. All subsequent DRAFT experiments will use 1024 for $d_{ada}$ since it is a good trade-off between performance and efficiency.

\subsection{Main Results for DRAFT}
\label{ssec:result_draft}

\begin{table*}[t]
\caption{WER results of SAFT and DRAFT for APC, Wav2vec2.0 and HuBERT on the OGI and MyST datasets. We do not provide baseline results for Wav2vec2.0 and HuBERT because we failed to obtain a reasonable WER without pretraining. \textbf{Finetune} is a system using pretraining but without adaptation, and \textbf{Adapter Finetune} is a system where only residual adapters are updated during finetuning. SAFT and DRAFT are systems with finetuning and an additional adaptation stage, and SAFT does not use residual adapters but DRAFT does. Bold numbers: the best results achieved and improvements are statistically significant. NC: no convergence.}
\footnotesize
\centering
\begin{tabular}{l  cc cc  cc cc  cc  cc }
\hline
\rule{0pt}{2ex}
\multirow{3}{*}{} & \multicolumn{4}{c}{APC} & \multicolumn{4}{c}{Wav2vec2.0} & \multicolumn{4}{c}{HuBERT} \\
\cmidrule(r){2-5} \cmidrule(r){6-9} \cmidrule(r){10-13}
\rule{0pt}{2ex}
~ & \multicolumn{2}{c}{OGI} & \multicolumn{2}{c}{MyST} & \multicolumn{2}{c}{OGI} & \multicolumn{2}{c}{MyST} & \multicolumn{2}{c}{OGI} & \multicolumn{2}{c}{MyST} \\
\cmidrule(r){2-3} \cmidrule(r){4-5} \cmidrule(r){6-7} \cmidrule(r){8-9} \cmidrule(r){10-11} \cmidrule(r){12-13}
~ & dev &  test &  dev & test & dev & test & dev & test & dev & test & dev & test \\
\hline
Baseline (w/o SSL) & 5.9 & 7.0 & 36.7 & 36.3 & - & - & - & - & - & - & - & -\\
Finetune & 5.0 & 6.1 & 32.2 & 31.6 & 2.27 & 2.70 & 17.84 & 17.16 & 2.07 & 2.48 & 17.40 & 16.71 \\
Adapter Finetune\cite{thomas2022efficient} & 8.6 & 10.1 & 47.4 & 47.3 & NC & NC & NC & NC & NC & NC & NC & NC  \\
\hline
\multicolumn{13}{c}{Self-Transfer} \\
\hline
SAFT & 5.0 & 5.9 & 33.4 & 32.9 & 2.22 & 2.67 & 17.85 & 17.28 & 2.02 & 2.43 & 17.52 & 16.89 \\
\textbf{DRAFT} & \textbf{4.4} & \textbf{4.9} & \textbf{30.1} & \textbf{29.4} & \textbf{2.11} & \textbf{2.51} & \textbf{17.21} & \textbf{16.70} & \textbf{1.85} & \textbf{2.05} & \textbf{16.79} & \textbf{16.53} \\
\hline
\multicolumn{13}{c}{Cross-Transfer} \\
\hline
SAFT & 5.2 & 6.2 & 37.8 & 37.3 & 2.33 & 2.85 & 21.24 & 20.28 & 2.11 & 2.30 & 17.67 & 17.20  \\
DRAFT & 4.7 & 5.5 & 31.4 & 30.8 & 2.13 & 2.63 & 17.95 & 17.36 & 2.03 & 2.28 & 17.13 & 16.65 \\
\hline
\end{tabular}
\label{tab:draft_all}
\end{table*}

We evaluate the DRAFT framework for three widely used SSL methods: APC, Wav2vec2.0 and HuBERT on both the OGI and MyST datasets. Results are shown in Table \ref{tab:draft_all}. As a comparison, we also include the results of adapter finetuning\cite{thomas2022efficient}, which uses residual adapters directly during the finetuning stage. Though it performs well in\cite{ thomas2022efficient}, adapter finetuning does not work well in our case. Specifically, the performance of adapter finetuning is much worse than that of finetuning the whole system. For Wav2vec2.0 and HuBERT, adapter finetuning does not even converge (100\% WER). We also compare DRAFT with SAFT (described in Sec.\ref{sssec:saft}). The table shows that SAFT yields a small improvement and sometimes a negative effect on the WERs compared to the finetuning baselines (without adaptation). The reason may be that updating the entire model causes a catastrophic forgetting of the knowledge learned from adult speech, even though we use a smaller learning rate to prevent overfitting for SAFT. However, when the proposed DRAFT framework is used, the WERs of the three SSL methods decrease for both the OGI and MyST datasets compared to the finetuning baselines (without adaptation). Specifically, we achieve relative WER improvements of 19.7\%/7.0\%, 7.4\%/2.7\%, and 16.0\%/1.1\% on the OGI/MyST test sets for APC, Wav2vec2.0, and HuBERT, respectively. The relative WER improvements with the OGI data are larger than those with the MyST data, which might be because SSL methods are more beneficial for a lower-resource task. Overall, HuBERT achieves the best WER for both datasets. Wav2vec2.0 and HuBERT are better than APC because they use non-causal transformers while APC uses a causal transformer. In addition, the number of APC parameters (39M) is smaller than that of Wav2vec2.0 and HuBERT (95M).


\subsection{Cross Knowledge Transfer of Residual Adapters}
We also examine the cross knowledge transfer ability of the residual adapters (RAs) learned from one dataset to another as they learn domain-related information. Specifically, the RAs trained with OGI data during the adaptation stage are used for training MyST data at the finetuning stage and vice versa. The results are shown in Table \ref{tab:draft_all}. Results show that when using the redisual adapters learned from MyST data, DRAFT can improve the WER performance on the OGI dataset consistently with the three SSL methods. This shows that DRAFT can be used for efficiently learning knowledge from one domain (child spontaneous speech), and then achieve better finetuning for another related domain (child read speech). However, the use of RAs learned from OGI data does not help the performance for MyST data. It might be because OGI has only 50-hour speech data, which is not large enough for adapting to the MyST data (240 hours). However, DRAFT always performs better than SAFT in cross knowledge transfer settings, which again shows DRAFT's ability at preventing catastrophic forgetting. We are also interested in determining whether the learned residual adapters from different domains can be fused together to further improve the performance. However, we do not observe improvements from preliminary results.

\section{Conclusions}
\label{sec:conclusion}

In this paper, we introduce self-supervised learning (SSL) methods for children's ASR when unannotated adult speech data are used in the pretraining stage. Specifically, a domain responsible adaptation and finetuning (DRAFT) framework was proposed to alleviate the domain shifting problem between pretraining (adult speech) and finetuning (child speech) data. The DRAFT framework performed well on APC (using causal transformers), Wav2vec2.0 and HuBERT methods (using non-causal transformers). When compared to the conventional finetuning baselines without adaptation, we achieved relative WER improvements of up to 19.7\% on the two child ASR tasks. The cross knowledge transfer experiments show a broader usage of the proposed DRAFT framework. To our knowledege, this and our previous papers\cite{fan2021bi,fan2022towards} are the first to develop SSL techniques for child ASR. A future direction of research could be to investigate the fusion of residual adapters learned from various domains for greater WER improvements on this and other low-resource ASR tasks.



\bibliographystyle{IEEEtran}
\bibliography{mybib}


\end{document}